# Complexity of the Online Distrust Ecosystem and its Evolution

**Lucia Illari[1,2], Nicholas J. Restrepo[3], Neil F. Johnson[1,2]***

[1]Dynamic Online Networks Laboratory, George Washington University, Washington, D.C., 20052, U.S.A.

[2]Physics Department, George Washington University, Washington, D.C., 20052, U.S.A.

[3]ClustrX LLC, Washington, D.C., U.S.A.

**\* Correspondence:**
Neil F. Johnson
neiljohnson@gwu.edu

**Keywords: betweenness centrality, complex adaptive systems, digital discourse dynamics, holistic strategies, network analysis, online misinformation, resilience, social media dynamics**

## Abstract

Collective human distrust – and its associated mis/disinformation – is one of the most complex phenomena of our time, given that approximately 70% of the global population is now online. Current examples include distrust of medical expertise, climate change science, democratic election outcomes -- and even distrust of fact-checked events in the current Israel-Hamas and Ukraine-Russia conflicts. Given that large social media platforms like Facebook have become pivotal battlefields, a new challenge for complexity science is to understand why the online distrust ecosystem is so resilient, how it evolves over time, and how well current mitigations and policies work. Here we adopt the perspective of the system being a complex dynamical network, in order to address these questions. We analyze a Facebook network of interconnected in-built communities (Facebook pages) totaling roughly 100 million users who, prior to the pandemic, were just focused on distrust of vaccines. Mapping out this dynamical network from 2019 to 2023, we show that it has quickly self-healed in the wake of Facebook's mitigation campaigns which include shutdowns. This confirms and extends our earlier finding that Facebook's ramp-ups during COVID-19 were ineffective (e.g. November 2020). We also show that the post-pandemic network has expanded its topics and has developed a dynamic interplay between global and local discourses across local and global geographic scales. Hence current interventions that target specific topics and geographical scales will be ineffective. Instead, our findings show that future interventions need to resonate across multiple topics and across multiple geographical scales. Unlike many recent studies, our findings do not rely on third-party black-box tools whose accuracy for rigorous scientific research is unproven, hence raising doubts about such studies' conclusions – nor is our network built using fleeting hyperlink mentions which have questionable relevance.

# 1    Introduction

There is arguably nothing more complex than human behavior. The COVID-19 pandemic showed how, in moments of uncertainty and concern, the global population (of which approximately 70% are now online) tend to go online to seek advice from their trusted online community (e.g. Facebook page) -- even though such communities do not typically have any documented medical or scientific expertise (Illari, L. 2022; Restrepo 2021; Restrepo 2022; Johnson et al. 2020). Somehow, the collective hearts and heads within and across these in-built communities get mixed in such a way that established medical and scientific expertise get set aside (Kleineberg and Boguñá, 2016; DiResta, 2018; Lappas et al., 2018; Semenov et al., 2019; Ardia et al., 2020; Chen et al., 2020; Starbird et al., 2020; Larson, 2020; Hootsuite, 2021; Douek, 2022; Rao et al., 2022; Getting Ahead of Misinformation, 2023; Johnson 2022). This can also leave people open to online manipulation through intentional disinformation which quickly expands to fill any void (Beyond disinformation – EU responses to the threat of foreign information manipulation, 2023; Nobel Prize Summit, n.d.; Catherine Meyers, n.d.; NewsDirect, n.d.; Zaid Jilani, n.d.).

This raises the crucial societal question going forward, of how a complex system of adaptive, decision-making humans with limited information, imperfect memories and knowledge bases, and often strong collective interactions, will behave in future global crises (San Miguel and Toral 2020; Ramirez et al. 2022; Czaplicka et al. 2022; Diaz-Diaz 2022). In particular, it opens up a new challenge opportunity for complexity science to understand why the online distrust ecosystem is so resilient, how it evolves over time, and how well do current mitigations and policies work at scale. We attempt to address these questions here.

We pursue a complex dynamical-network systems approach to understanding this problem (Mazzeo and Rapisarda 2022; Mazzeo et al. 2021; Tria et al. 2018; Yalcin et al. 2016) in contrast with the rather static existing health approach where the focus is either on the entire population (public health) or a single individual (personal health). Instead, we focus in on the intermediate meso-scale dynamics between the scale of a single individual and the total population, because it is known from complex systems studies that this is the dynamical regime where correlations appear that can then lead to macroscopic changes and transitions. This mesoscopic focus has a direct real-world justification because the online human social media system contains approximately 5 billion people who self-organize into in-built communities – each of which contains many thousands of individuals and has nothing to do with network community detection (Illari, L. 2022; Restrepo 2021; Restrepo 2022; Johnson et al. 2020). Each in-built community is a separate Facebook page in our study. These communities (Facebook pages) then interconnect with each other via page-level links in order to share content feeds and hence share conversations, opinions and information at the page level. Reciting the description of a complex system from Ref. (Johnson et al. 2003) and elsewhere, we expect the online human system to feature the key complex systems characteristics: (i) Feedback which can change with time -- for example, becoming positive one moment and negative the next -- and may also change in magnitude and importance. (ii) Non-stationarity. We cannot assume that the dynamical or statistical properties observed in the system's past, will remain unchanged in the system's future. (iii) Many interacting agents. The system contains many individuals or 'agents', which interact in possibly time-dependent ways. Their community-level behavior will respond to the feedback of information from the system as a whole and/or from other communities. (iv) Adaptation. A community can adapt its behavior to reflect world events and attract new recruits. (v) Evolution. The entire online population evolves, driven by an ecology of interacting communities which interact and adapt under the influence of feedback. The system typically remains far from equilibrium, and hence can exhibit extreme behaviors. (vi) Single realization. The system under study is a single realization, implying that standard techniques whereby averages over time are equated to averages over ensembles, may not work. (vii) Open system. The system is coupled to the outside world including local and world news, and hence responds to both exogenous (i.e. outside) and endogenous (i.e. internal, self-generated) effects.

As we show in this paper, adopting this complex dynamical-network systems perspective yields various new insights – in particular, why the distrust ecosystem has become so resilient, how it evolves over time, and how well current mitigations and policies work. The important task of providing an accompanying mathematical description is a work in progress which could benefit from fascinating recent analyses (see for example San Miguel and Toral 2020; Ramirez et al. 2022; Czaplicka et al. 2022; Diaz-Diaz 2022; Mazzeo and Rapisarda 2022; Mazzeo et al. 2021; Tria et al. 2018; Yalcin et al. 2016). While we ourselves cannot yet offer such a mathematical theory, we hope that the empirical patterns presented here will stimulate such work in the complex systems community. With this in mind, our analysis here focuses around establishing some 'stylized facts' using several complex systems tools in order to help future model-building – specifically, we use dynamical networks and n-Venn diagrams for our analysis.

There is of course already a vast literature outside the complex systems field, attempting to understand the problem -- mostly from sociological, psychological, political science, and pure computer science perspectives. In addition, the social media giants such as Facebook (Meta, but we refer to them as Facebook for convenience) have made efforts to curtail the associated avalanche of mis/disinformation through strategies like content moderation and removal of anti-vaccine content. We refer for example to Van der Linden et al., 2017; Lazer et al., 2018; Lewandowsky, Stephan et al., 2020; Rory Smith et al., 2020; VERIFIED: UN launches new global initiative to combat misinformation, 2020; Getting Ahead of Misinformation, 2023; Restoring Trust in Public Health, 2023; Get the facts on coronavirus, n.d.; Government to relaunch 'Don't Feed the Beast' campaign to tackle Covid-19 misinformation – Society of Editors, n.d.; Navigating Infodemics and Building Trust during Public Health Emergencies A Workshop, National Academies, n.d.; New USD10 Million Project Launched To Combat the Growing Mis- and Disinformation Crisis in Public Health, n.d.; Understanding and Addressing Misinformation About Science A Public Workshop, National Academies, n.d.; What's Being Done to Fight Disinformation Online, n.d.; Calleja et al., 2021; Roozenbeek et al., 2022; Mirza et al., 2023; Wanless, n.d.; WHO, n.d. However, distrust continues to persist (Ravenelle et al., 2021; Boyle, 2022) which suggests that these interventions have only scratched the surface. Here we pursue the understanding of why this is, in order to gain insight into what is missing from current efforts and hence what can be done.

Our specific contribution here -- which is made possible only because we take a complex systems perspective – is the first ever analysis of the post-pandemic evolution of Facebook's vaccine discourse network through September 2023. This enables us to chart the trajectory that the network has taken from its initial November 2019 state when it was focused around vaccine distrust (as reported in Johnson et al., 2020). Our findings reveal a vast web-of-distrust that has developed post-pandemic, that seamlessly integrates diverse topics, locations, and geographical scales and hence challenges traditional approaches to health messaging. Our analysis focuses on Facebook because of its unparalleled reach as the predominant social media platform (Meta, 2020). Central to our study are its inherent community features—Facebook Pages—that facilitate virtual congregations centered on shared interests, such as parenting. These digital communities have emerged as pivotal hubs for social interactions on platforms like Facebook. Contemporary research highlights the growing dependence on these communities for guidance, particularly in areas like family health (Ammari and Schoenebeck, 2016; Laws et al., 2019; Moon et al., 2019). Members develop a profound trust within their respective communities, often valuing insights from peers with parallel concerns. For instance, parents navigating health decisions for their children tend to resonate more with anecdotal experiences and perspectives shared by fellow parents (Ammari and Schoenebeck, 2016; Laws et al., 2019; Moon et al., 2019).

Our analysis in this paper yields three major insights about the nature of this post-pandemic Facebook vaccine discourse network: (1) an inherent resilience within the network, allowing it to adapt and counteract Facebook's interventions; (2) the recent convergence of diverse narratives, from COVID-19 to climate change and elections, into a unified web of discourse, underscoring the expansive reach and influence of the network; and (3) the switching of discourse dynamics between global and local scales, demonstrating the network's versatility and adaptability. This means that traditional mitigation strategies are insufficient for improving distrust at scale because they tend to operate in silos for a specific topic (e.g. monkeypox) and at a specific geographical scale (e.g. state-level health departments for the individual states in the U.S.). It also enables us to propose a potentially far better system-level approach in the form of interventions that resonate across multiple topics and across multiple geographical scales. Our findings also serve to confirm and extend our earlier finding reported in late 2021 and early 2022, that Facebook's ramp-ups were ineffective during COVID-19 (e.g. ramp-up around November 2020) as shown in Refs. (Restrepo 2022, Johnson 2022). This lack of impact of Facebook's mitigation policies during the pandemic (e.g. ramp-up around November 2020) was subsequently independently confirmed in a later separate study (Broniatowski 2023b).

## 2    Materials and Methods

We examine how the distrust discourse changed post-pandemic by analyzing the evolution through time of the pre-pandemic 2019 Facebook network that had developed around vaccine distrust (Johnson et al., 2020). It comprised 1356 interlinked communities that were pro-vaccine, anti-vaccine, or neutral towards vaccines. Our methodology follows exactly Ref. (Johnson et al., 2020) and the details are reviewed again in the Supplementary Material (SM). Each Facebook page is a community (i.e., node in Fig. 1) with a unique ID. These communities have nothing to do with community detection in networks. These communities provide spaces for users to gather around shared interests, promoting trust within the community and potential distrust of external issues (Kim and Ahmad, 2013; Forsyth, 2014; Chen et al., 2015b, 2015a; Gelfand et al., 2017; Ardia et al., 2020; Arriagada and Ibáñez, 2020; Madhusoodanan, 2022; Pertwee et al., 2022; Tram et al., 2022; Freiling et al., 2023; Song et al., 2023; WHO, n.d.). Our trained researchers manually and independently classified each community as pro-vaccine, anti-vaccine, or neutral. In cases of disagreement, the researchers discussed until reaching consensus. This process yielded a network of 1356 interlinked communities across countries and languages, with 86.7 million individuals in the largest network component classified as follows: 211 pro-vaccine communities with 13.0 million individuals (blue nodes, Fig. 1); 501 anti-vaccine communities with 7.5 million individuals (red nodes, Fig. 1); and 644 neutral communities with 66.2 million individuals (non-blue or red nodes, Fig. 1). The neutral communities were further sub-categorized by type, such as parenting.

To categorize the discourse topics within each of the 1356 communities, we developed keyword filters for five non-vaccine-related topics that subsequently emerged: COVID-19, monkeypox (mpox), abortion, elections, and climate change. The filters combined regular expressions and keyword searches in multiple languages (details in the Supplementary Material (SM) Section 7) and were applied to post content, descriptions, image tags, and link text. By 'non-vaccine' we mean topics not centered on vaccines in a broad sense. Although the COVID-19 and monkeypox filters included some vaccine terms (for e.g., "monkeepox vax'n"), the intent was to capture disease-specific discourse rather than general vaccine discussion. Distinguishing between COVID-19 and monkeypox posts in an automated manner would have been difficult without additional human classification had we been filtering for general vaccine discussion rather than disease-specific discourse.

A link between communities *A* and *B* (i.e. Facebook pages *A* and *B*) indicates community (page) *A* recommends community (page) *B* to its members at the page level, creating a hyperlink from page *A* to page *B*. This shows page *A*'s interest in page *B*'s content and goes beyond just mentioning *B* casually in a URL. Links do not necessarily signify agreement, rather it directs the attention of *A*'s members to *B*, and vice versa it exposes *A* to feedback and content from *B*. While not all members will necessarily pay attention, a committed minority of 25% can be enough to influence the stance of an online community (Centola et al., 2018).

Of the 1356 communities, 342 are local with 3.1 million individuals. The remaining 1014 are global with 83.7 million individuals. In terms of geography, a global community is a page with broad, worldwide focus that is not tied to a specific location, while a local community is focused on a specific geographic area, such as a neighborhood, city, county, state, or country (e.g., "Vaccine information for parents" or "Global Trends" vs "Vaccine information for Los Angeles County parents"). In terms of topic, a global community discusses diverse issues broadly, whereas a local community has a narrow topical focus (e.g., pages discussing only elections). The size of each community can be estimated by the number of likes, given that the average user only likes one Facebook page on average (Hootsuite, 2021). However, our analysis and findings are not dependent on this.

In this paper, the terms 'global' and 'local' are applied to two categorical dimensions—geographic scale and topic—in order to examine the interconnection between geographic and topic glocality. This dual usage of 'global' and 'local' elucidates how geographic and topic glocality can interrelate and allow communities to take up different glocal positions. For instance, a community focused on a narrow topic within a small geographical locality embodies hyperlocality on both dimensions, while one that discusses many topics worldwide embodies hyperglobality. The purpose of this is to analyse the extent to which the post-pandemic, skepticism discourse—whether it be about global health challenges, societal decisions, or environmental concerns—expanded beyond just hyperlocal geography and topics to encompass more hyperglobal geography and topics. The terms 'global' and 'local' effectively convey this evolution across both dimensions.

We use betweenness centrality to quantify nodes' potential to serve as an information conduit. A node with high betweenness centrality is critical to information flow and is considered a 'broker' of information, which is instrumental in understanding the spread of vaccine-related narratives. This quantity is discussed in greater detail in all network science textbooks.

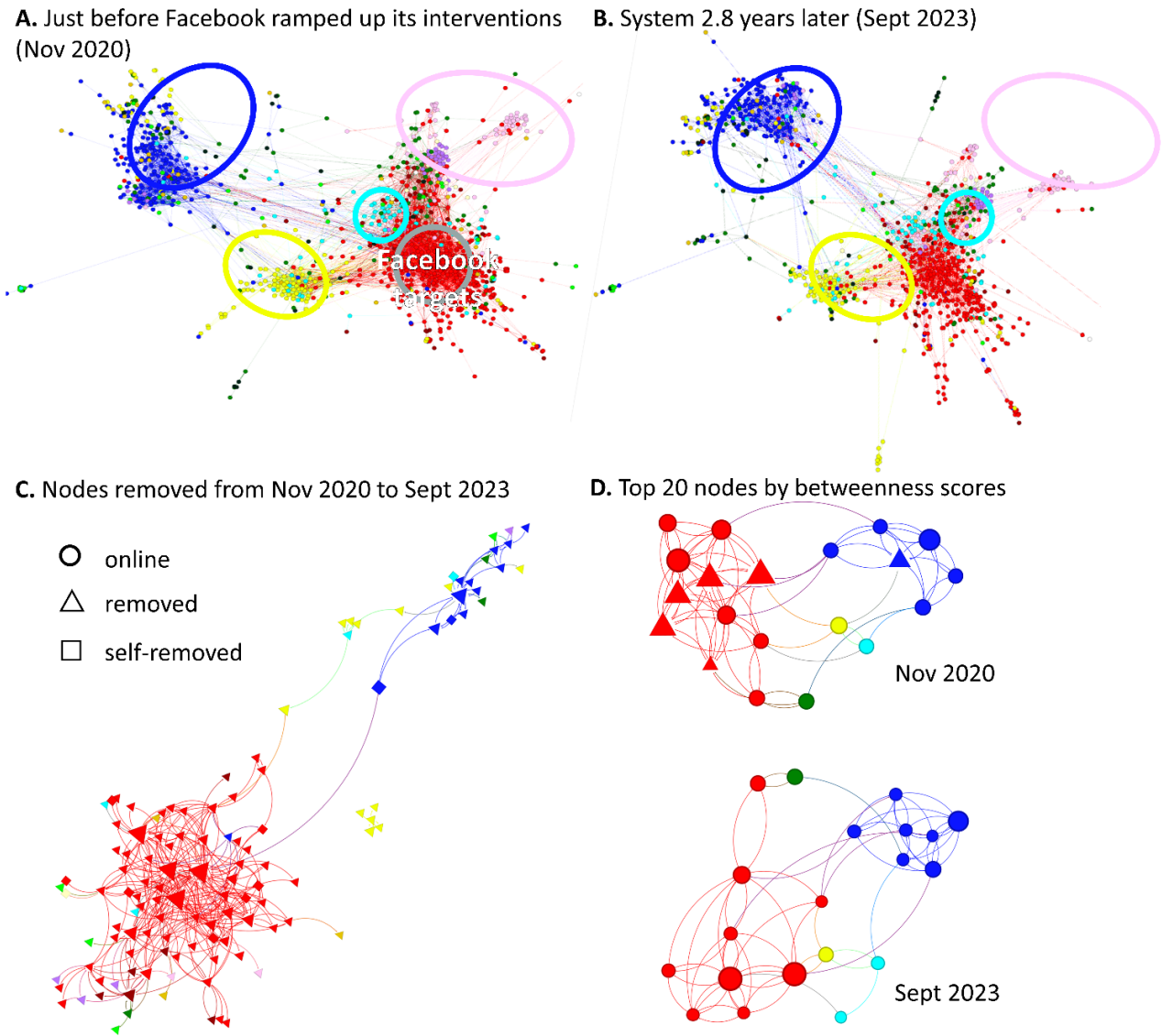

**FIGURE 1.** Evolution over time of the 2019 Facebook vaccine distrust network (Johnson et al., 2020) which encompasses nearly 100 million users. Starting in November 2020, Facebook amplified its interventions, notably removals (see (C)). Until that point (i.e., (A)), the focus was on posting factual messages. The ForceAtlas2 layout mechanism means that nodes (i.e., Pages) appearing closer together are more likely to share content. (A) and (B) use the same scale: rings indicate categories of neutral Pages maintaining the same position, underscoring the intensification of connections within the red (i.e. anti-vaccine) nodes and between red and neutral nodes. Blue nodes are pro-vaccine communities. Different colors are used to distinguish the specific categories of the neutral communities (i.e. the 'greens'). (D) reveals a self-repairing core 'mesh' of high betweenness anti-vaccination (red) communities from within (A)(B) that can—and does persist to—share and distribute extreme content amongst each other and with other nodes in the full ecology, including millions of mainstream users in neutral nodes. The specific nodes within this core mesh (i.e., (D) lower panel) change over time, but its structure remains resilient which serves as confirmation of the inefficiency in Facebook's November 2020 (and other) interventions as reported in Ref. (Restrepo 2022). This lack of impact was subsequently confirmed in a later study (Broniatowski 2023b). Node size in (C)(D) signifies betweenness values, i.e., a node's capacity to serve as a conduit for content.

Distinct methodological advantages of our approach as compared to other recent studies (Broniatowski et al. 2023a, 2023b) include the fact that our findings do not rely on third-party black-box tools such as CrowdTangle whose accuracy for rigorous scientific research is unproven and increasingly questionable given Facebook's internal policies, hence casting doubt on these other studies' conclusions. Also, these previous studies are restricted since they only consider a mention/hyperlink network where links represent the appearance of a URL in a page's content. There is no a priori reason or guarantee that these fleeting URLs represent meaningful long-term connections between one community and another, nor that they influence their subsequent behavior in any significant way. This needs to be proven before any subsequent network analysis can be regarded as meaningful at scale. By contrast, in our study the links between nodes (communities) are better defined, i.e. if page A 'likes' page B then this community A links to community B which creates an information conduit from B into A and hence exposes A's users to B's content (e.g. new posts from Facebook page B can appear on Facebook page A). So the difference between our network versus those of these other studies, lies in the fact that our study is a follower/like/social network based on the formal page follow/like connections on Facebook itself whereas those other studies are simply a mention network. Technically, we also note that it is not obvious how to create our kind of network using third-party tools such as CrowdTangle as used by these other studies.

## 3    Results

Using the pre-COVID-19 vaccine network from Ref. (Johnson et al., 2020) as a starting point, we mapped the evolving dynamics of these interconnected Facebook communities (each community is a Facebook Page) and their views (Fig. 1(A)-(D)). Around November 2020, Facebook shifted its approach from primarily posting informational messages to aggressively removing or suppressing content, especially from anti- communities that shared vaccine-skeptical content. However, while these interventions were aimed at curbing the spread of misinformation, our results suggest that the more subtle mainstream science-skeptical content, which may not be outright false but still casts doubt, continued to permeate the network, reaching a broader audience. Figs. 1A and B chronicle the evolution of the 2019 Facebook vaccine-view ecology. Fig. 1(A) captures the ecosystem just before Facebook intensified its interventions in November 2020, while Fig. 1(B) offers a snapshot nearly three years later (the network at more timestamps is available in SM Section 2). The force-directed layout, ForceAtlas2, implies that nodes that appear nearby to each other in the network layout (i.e., communities/Pages) are more likely to exchange content. The concentric rings in both figures demarcate the original positions of the communities from the pre-COVID-19 era of the network (end of 2019), emphasizing the strengthening connections within and between different nodes, particularly within the red nodes and between the red and neutral nodes.

Fig. 1C illustrates nodes that were either deleted or made private between November 2020 and September 2023. Here, node size is related to betweenness scores and fan count/follower count for these pages range from millions to tens. Notably, despite the removal of some large, active extreme communities, the network's high interconnectivity and operational resilience remains largely intact. This resilience is further highlighted in Fig. 1(D), which depicts the top 20 nodes based on betweenness centrality—an indicator of the influence a node has over the flow of information in a network. Most notably, red nodes, signifying anti-vaccination communities, rank prominently within this subsystem of top nodes. Even though numerous red nodes underwent removal by Facebook (as seen in Fig. 1(D) upper panel, as indicated by the triangular nodes), a network rewiring phenomenon ensued, leading to the emergence of other red nodes in their stead (shown in Fig. 1(D) lower panel). A similar resilience effect can be observed between November 2019 and November 2020 (see SM Section 2). This confirms our earlier report of the inefficiency in Facebook's November 2020 (and other) interventions (Restrepo 2022; Johnson 2022). This initial finding (Restrepo 2022; Johnson 2022) was subsequently confirmed by a later independent study (Broniatowski 2023b).

From the outset, the network experienced an overall upward trajectory in betweenness scores. The magnitude of this elevation varies among top nodes, ranging from a percent difference of +7.98827% to +139.997%. This variation, as captured by a mean absolute deviation of 20.9697%, underscores the diverse trajectories of nodes within the network. Despite periodic changes and distinct node shifts between November 2020 and September 2023, the vaccine stance composition of this group exhibits remarkable consistency. This stability extends beyond mere node categorization, as can be seen in Figs. 1(A)(B) — i.e. the holistic structure and dynamics of the network have remained largely consistent.

The resilience observed in the network's core structure, even in the face of external interventions, has profound implications. It indicates that there exists a robust and interconnected subset of nodes that consistently rise to prominence, effectively preserving the network's thematic and structural integrity. This resilient core seems to be somewhat impervious to node deletion mitigation efforts, suggesting that such interventions alone might not suffice to significantly alter the discourse or dynamics within the network. Instead, the network appears to have an inherent capacity to 'self-heal' and maintain its equilibrium. In essence, while individual nodes may come and go, the network's backbone remains steadfast. This resilience underscores the challenge of influencing or shifting discourse dynamics solely through node-centric interventions. A more holistic, system-level understanding, and approach is therefore necessary to effectively navigate and influence such resilient networks.

Delving deeper into the geographic specificity and thematic diversity of Facebook communities with Fig. 2(A), we observe that the of the top 20 nodes by betweenness, only 4 are geographically localized. This implies that while most top communities cater to a broader audience, there's a significant subset that zeroes in on local concerns and narratives. While global narratives set the overarching tone, it's often the local narratives that drive grassroots actions, sentiments, and perceptions. Thus, their presence underlines the importance of region-specific dialogues even within a global platform like Facebook, and suggests that localized information, when tailored effectively, can have a disproportionately large impact.

Furthermore, a notable 70% of these top nodes engage in dialogues spanning five key discourses: COVID-19, mpox, abortions, elections, and climate change. The depth of their engagement, however, is not uniform. A minority (3 of 14 nodes) primarily focus on one topic, while the majority (11 of 14 nodes) partake in discussions on at least three topics. The most recurrent thematic combinations are COVID-19, mpox, and elections, followed by COVID-19, mpox, and climate change. This suggests that these communities are not siloed into singular narratives but rather weave together multiple themes, potentially drawing parallels, intersections, or contrasts between them.

Statistical analysis helps us further dissect the relationships between a page's fan count, its betweenness centrality, and the breadth of topics with which it engages. While there is not a significant relationship between fan count and betweenness centrality, our analysis reveals two key findings: (1) Pages discussing a wider range of topics exhibited a weak, yet statistically significant, positive correlation with betweenness centrality ($r = 0.121$, $p < 0.001$). This suggests such pages may play a more central role in information dissemination within the network. (2) Larger audience pages (higher fan count) show a propensity to discuss a diverse set of topics, as evidenced by the weak positive correlation between fan count and topic count ($r = 0.121$, $p < 0.001$). Furthermore, one-way analysis of variance highlighted significant differences in both betweenness centrality scores ($F(5, 1093) = 4.14$, $p < 0.001$) and fan counts ($F(5, 1093) = 4.53$, $p < 0.001$) across different topic count categories.

This means that a large follower count does not necessarily equate to a page being a central hub for information flow within the network. Betweenness centrality, which gauges a node's role as a connector in the network, reveals that a page's influence is not solely dictated by its fan base. For example, a celebrity page with many followers may not be a primary information source, whereas a niche expert page might have fewer followers but serves as a pivotal connection point in certain discussions. Interestingly, while a page's audience size does not correlate with its network centrality, the breadth of topics that it covers does. A diverse audience will have varied interests, pushing the page admins to cover multiple subjects to keep the audience engaged. Essentially, pages that delve into various topics not only position themselves more centrally in the network but also appeal to a broader audience, highlighting the importance of content diversity in influencing both network prominence and audience engagement.

Fig. 2(B) delves into the geographic indications within community names. Predominantly, these point to the United States. However, the top 10 also features other nations, primarily English-speaking, such as Australia, Canada, and the United Kingdom. Additionally, while page administrative data included many of these same countries for their administrators, it showcases diversity with countries such as Israel, the Philippines, and Slovenia, which are not represented in page usernames. This suggests strategic branding: pages may adopt names that resonate with specific audiences, even as they maintain a globally diverse operational team. As previously mentioned, only 4 of the top 20 communities, when ranked by betweenness centrality, incorporate explicit geographic references in their usernames. This absence might be a conscious choice, either to appeal to a global audience or to transcend regional confines to address broader themes. The involvement of moderators from non-English speaking countries further emphasizes this global engagement, hinting that the discussions resonate universally, transcending the initial impressions given by geolocalized network indicators. In essence, while some Facebook communities may appear geographically or thematically localized, the overarching network intertwines global and local, highlighting the nuanced nature of discourse on the platform.

**A.** Full system in Sept 2023

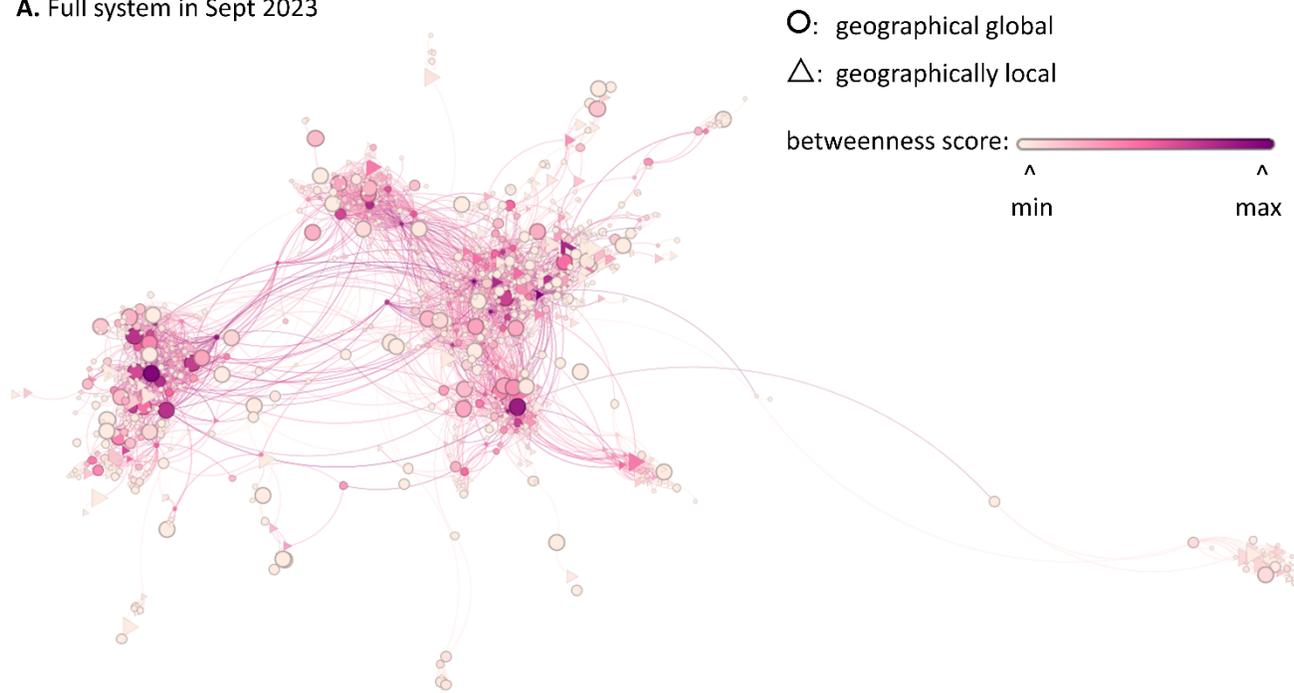

**B.** Geolocalized network

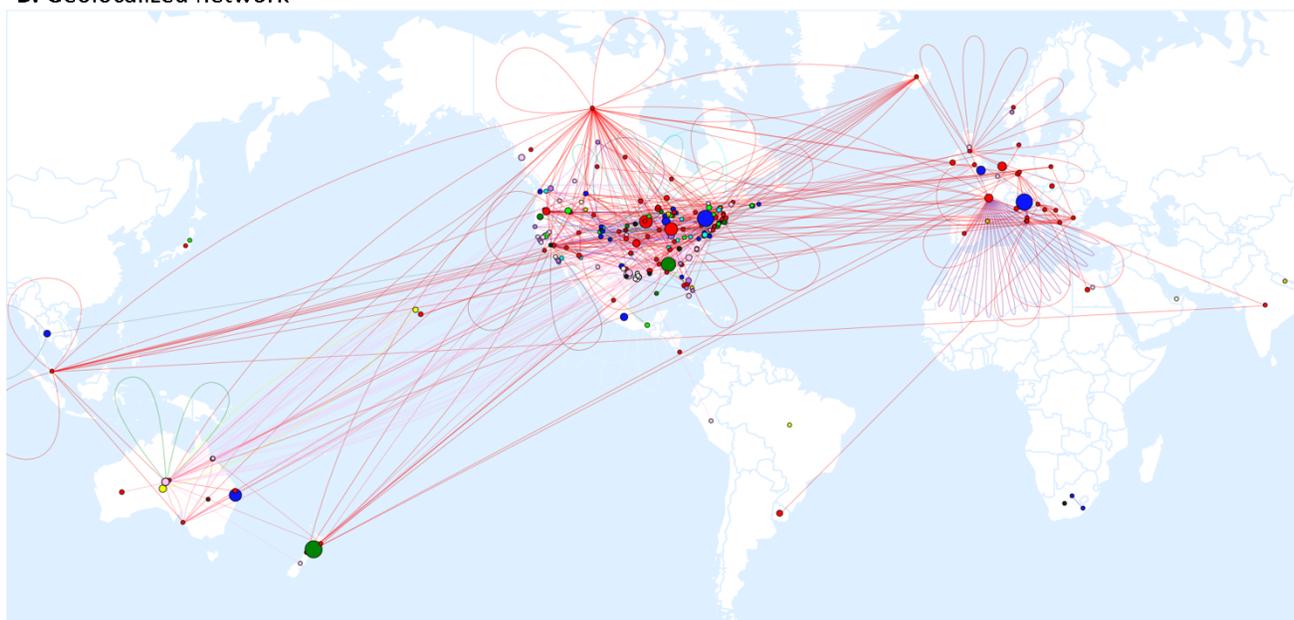

**FIGURE 2.** Visualization of the post-pandemic 2019 Facebook vaccine network. (A) Nodes (communities) are sized by number of discourse topics in which each engaged (topic count), shaped by local or global status (whether or not a location is mentioned in a Page's username), and colored by betweenness centrality score. Visual examination suggests a positive correlation between topic count and betweenness score. (B) Geolocalized network of communities mentioning locations in usernames. Self-loops are not possible; the apparent 'self-loops' are due to the extremely close proximity of locations mentioned in community usernames. A clear majority of local nodes appear to be in English-speaking countries.

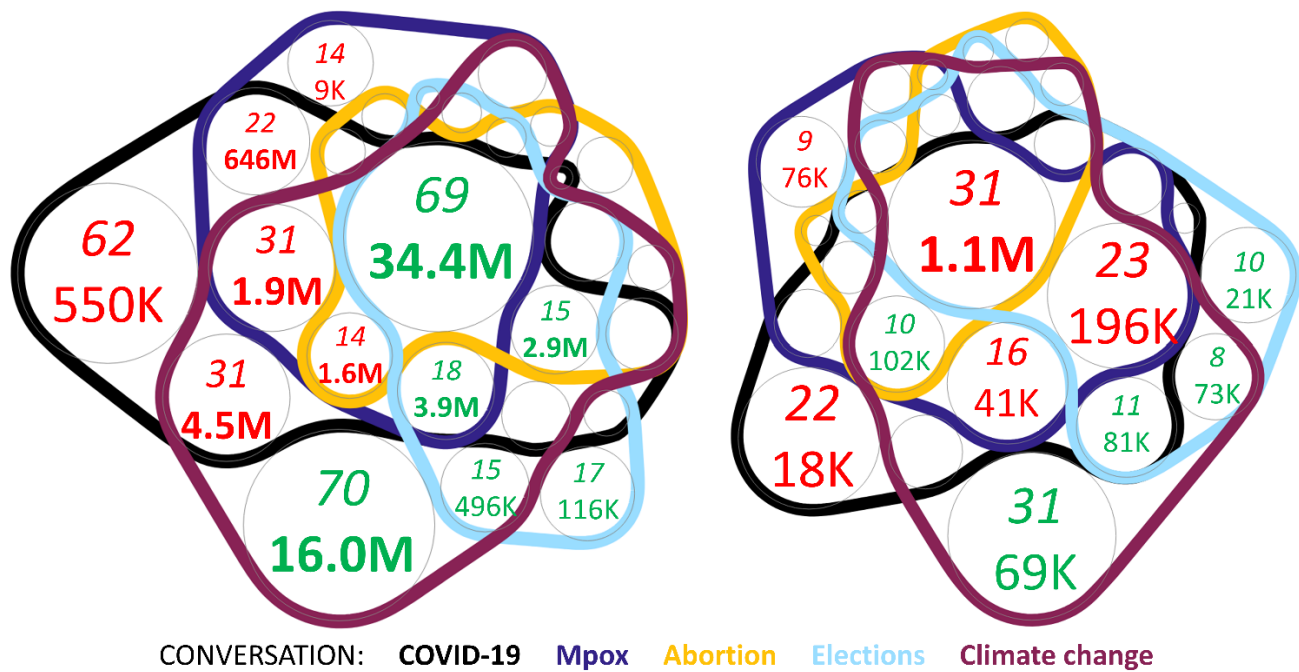

**FIGURE 3.** Topic mix within (A) global communities, and (B) local communities, shown using n-Venn diagrams (see SM Sect. 8 for simple example explaining an n-Venn diagram (Pérez-Silva et al., 2018)). Number of communities in italics on top, with number of individuals on bottom (bolded if >1M). Number is shown in red (or green) if there is a prevalence of anti-vaccination (or neutral) communities. Only regions with >3% of total communities have labels.

Figs. 3(A) and 3(B) highlight the significance of topic combinations in shaping public messaging. In total, an audience of 77.3 million individuals has engaged with one or more of these topics. The group of communities discussing all five topics ranks as the second largest, reaching a staggering 35.5 million individuals. Neutral communities primarily drive this discourse, however in their absence, anti-vaccination communities would dominate the narrative. Out of 29 potential topic combinations, only three would be led by pro-vaccination communities without the influence of neutral voices.

A noticeable trend is the limited number of communities focusing solely on topics like mpox, abortion, or elections. Meanwhile, discussions often intersect between COVID-19 and climate change, especially in global communities. For locally oriented communities, the pairing of elections and climate change emerges as a combination. Despite the participation of pro-vaccination communities, the discourse is frequently steered by communities whose perspectives diverge from the prevailing scientific consensus—and in many cases, these are communities that actively oppose it, especially at the local level. A case in point: while global communities discussing all five topics are majority neutral, their local counterparts are majority anti-vaccination. Moreover, if neutral communities were excluded, six topic combinations for global communities would be led by pro-vaccination voices. In contrast, this number shrinks to merely three for local communities.

Overall, our updated analysis of the original 2019 Facebook vaccine-view network reveals intricate patterns of interconnectedness, resilience, and thematic diversity. As global and local narratives intertwine, it emphasizes the platform's complex web of voices, underscoring the importance of understanding online discourse at scale as a complex system and hence its broader implications.

## 4    Discussion

Our analysis of the evolving dynamics within the Facebook community network paints a vivid picture of the intricate and resilient nature of online discourse, particularly concerning the 5 topics of COVID-19, mpox, abortions, elections, and climate change. The results underscore the network's adaptive capacity, a trait that conventional mitigation strategies might struggle to address effectively.

The resilience observed in the network's core structure, even amid external interventions, suggests a robust system that challenges traditional approaches to information dissemination and correction. Such resilience is emblematic of ecosystems that possess an inherent ability to maintain equilibrium, even when certain nodes are altered or removed. This resilience implies that singular, node-centric interventions, like the removal of certain Facebook communities, may only offer short-term solutions. For lasting impact, a broader, system-level approach is imperative. Specifically, our study reveals the multifaceted, interconnected structure of the online discourse on Facebook around topics that generate mis/disinformation and distrust. Its system-wide interconnected structure and resilience – despite Facebook's many interventions and mitigation policies, particularly during COVID-19, e.g. around November 2020 -- suggest that traditional, isolated mitigation strategies to address mis/disinformation and distrust of medical expertise across Facebook may fall short. These findings also serve to confirm and extend our analyses that appeared in 2022 which showed that Facebook's mitigations and policies during COVID-19 were inefficient in terms of disrupting this network of distrust (Restrepo et al. 2022; Johnson et al. 2022; Restrepo et al. 2021; Illari et al. 2022). This earlier finding (Restrepo 2022; Johnson 2022) was subsequently confirmed by a later independent study (Broniatowski 2023b).

A significant takeaway from our research is the importance of geographic and thematic diversity in shaping online narratives. While global themes dominate the overarching discourse, local narratives play a crucial role in driving grassroots sentiments and actions. This 'glocal' blend of global and local perspectives hints at the need for more nuanced and diversified messaging strategies. Rather than narrowly focusing on singular topics or scales, a combination of strategic topics, both local and global, might be more effective in addressing misinformation and fostering constructive discourse.

Furthermore, our analysis of topic combinations in shaping public messaging provides invaluable insights into the dynamics of online discussions. The intertwining of discussions, such as those between COVID-19 and climate change, suggests that communities aren't confined to siloed narratives. Instead, they weave multiple themes, drawing parallels or contrasts, a trait that mis/disinformation mitigation strategies should consider.

While our dataset is robust, it is worth noting that it primarily captures select languages prevalent on Facebook. However, the diverse locations of page administrators suggest a broader geographic representation. Future research could delve deeper into sentiment analysis or employ natural language processing techniques for richer insights. Moreover, while Facebook is a significant platform, understanding discourse dynamics on other social platforms would provide a more holistic view of the online vaccine distrust ecosystem.

Many other studies being published on similar themes, use CrowdTangle which is a commercial application tool owned by Facebook, and hence have claims and findings that rely entirely on the hope that CrowdTangle is accurate (Broniatowski et al. 2023a, 2023b). However, the researchers in such studies sit outside Facebook's CrowdTangle team and hence have little knowledge or quantitative explanation of how and why this tool returns the results that it does. In other words, it is effectively a black-box tool, which makes it unacceptable for academic science research in our opinion. Worse still, it is widely reported that the CrowdTangle tool is no longer being properly maintained and hence any accuracy that it may arguably have had in the past has now likely degraded significantly. It may still be the best tool available, but the lack of such tools means that simply being best has no bearing on whether it is sufficiently reliable and accurate for scientific research – and hence whether the often subtle conclusions from such studies carry any weight. We further expand on this in the SM (see also Restrepo et al. 2021 for our original critique). In addition, we stress that our study builds and analyses a follower/like/social network based on the formal page follow/like connections on Facebook itself, whereas those other studies rely on a mention network in which the links are fleeting URLs of unknown relevance or importance at the scale of the community. Technically, we also note that it is not obvious how to create our kind of network using third-party tools such as CrowdTangle as used by these other studies.

Finally, we comment on how our study's findings might help solve the problem of online distrust and its associated mis/disinformation without having to rely on social media platforms to finance costly human mitigation teams. It involves the novel idea proposed and later tested and presented by Restrepo et al. (see Restrepo et al. 2022 and earlier online public talks). In this scheme, the complex dynamical network of interlinked communities that is presented in this current paper, can be used to guide the formation of small anonymous deliberation groups – with the remarkable result that softening of the extremes within such groups is observed to arise in the empirical studies (Restrepo et al. 2022). Specifically, the extremes of the deliberation group get softened organically as long as the group contains individuals drawn fairly evenly from across the anti-, neutral and pro- spectrum. Moreover, this scheme could easily be automated at scale across social media platforms in a rapid and cost-effective way – which allows platforms to avoid engaging in top-down removal of communities, or providing preventative messaging, or censorship. We refer to Ref. (Restrepo et al. 2022) for full details, empirical and theoretical validation, and the core social science theory underpinning it. Hence the cost to social media platforms will be minimal – just the time cost of inviting individuals into new Facebook groups, for example, from the anti, neutral and pro communities. Importantly, this cost can be a minimal one since this process can be completely automated given the network map that we provide in this paper. We note that this remarkable effect appears to have been re-discovered and hence verified to some degree in a subsequent independent study by Abroms et al. in which Facebook groups effectively act as online deliberation groups (Abroms 2023).

In conclusion, our study offers a complex systems understanding of the Facebook distrust ecosystem and its post-pandemic shifts. Our findings underline the need to rethink current mitigation strategies, advocating instead for a more holistic approach that considers the network's diverse and interconnected nature. Harnessing these insights could pave the way for more effective interventions.

## 5    Conflict of Interest

The authors have no competing interests, either financial and/or non-financial, in relation to the work described in this paper.

## 6    Author Contributions

L.I. analysed the results and generated the figures. N.J.R. obtained and classified the data. N.F.J. supervised the project. L.I and N.F.J. wrote the paper. All authors were involved in reviewing the final manuscript, and in the conceptualization, methodology, and validation.

## 7    Funding

N.F.J. is supported by U.S. Air Force Office of Scientific Research awards FA9550-20-1-0382 and FA9550-20-1-0383, and by the Templeton Foundation in collaboration with S. Gavrilets.

## 8    Supplementary Material

Supplementary Material (SM) contents:

Section 1:    Methodology: Collecting data, building network, use of term "glocal", and analysis in the paper.

Section 2:    Breakdown of page admin country locations. Color scheme for neutral nodes in network plots. The system before COVID-19, one year later, and three years later. Geolocalized network. Preventing node overlap in Fig. 1 networks.

Section 3:    Classification of neutral nodes.

Section 4:    Example of Facebook banners promoting best-science Covid-19 guidance. Positions in network of the nodes that receive Facebook banners promoting best-science Covid-19 guidance.

Section 5:    ForceAtlas2 layout and analysis showing dependence of layout on strength of bonding.

Section 6:    Topic filter. System filtered by topic. System in October 2022 without node labels, and only with those labels appearing in Fig. 1. Fig. 1C–G in the main paper, with node labels.

Section 7:    Chi-square test for topic-glocality. Statistical results for fan count, topic count, and betweenness centrality scores.

Section 8:    Use of nVenn diagrams. General system dynamics, glocal breakdown, and comparison to simulation.

Section 9:    Relationship between the number of topics and glocality. Topic heatmaps.

## 9    Data Availability Statement

All data needed to evaluate the conclusions in the paper are present in the paper and Supplementary Material. The code used to generate Figs. 1 and 2(A) is Gephi, which is free open-source software. Figure 2(B) was generated using Mathematica, and Fig. 3 was generated using the nVennR package for R (vqf, 2023).